\documentclass[10pt,conference]{IEEEtran}
\IEEEoverridecommandlockouts
\usepackage{amssymb}
\usepackage{amsfonts}
\usepackage{amsmath}
\usepackage{tikz, pgfplots}
\usepgfplotslibrary{groupplots}
\pgfplotsset{width=\textwidth*0.4,compat=1.9}
\usepackage{siunitx}
\usepackage{svg}
\usetikzlibrary{positioning}
\usepackage{url}

\usepackage{algpseudocode}
\usepackage{algorithm}
\usepackage[libertine]{newtxmath}
\usepackage{subfig}
\usepackage[hidelinks]{hyperref}
\usepackage{orcidlink}
\def\BibTeX{{\rm B\kern-.05em{\sc i\kern-.025em b}\kern-.08em
    T\kern-.1667em\lower.7ex\hbox{E}\kern-.125emX}}
\makeatletter
\def\endthebibliography{%
  \def\@noitemerr{\@latex@warning{Empty `thebibliography' environment}}%
  \endlist
}

\usepackage[nolist]{acronym}

\ifCLASSINFOpdf
\else
\fi

\begin{document}
\title{A Deep Reinforcement Learning-based Approach for Adaptive Handover Protocols in Mobile Networks}
\author{\IEEEauthorblockN{Peter J. Gu\, \orcidlink{0009-0002-7853-9201}, Johannes Voigt\, \orcidlink{0000-0002-4032-5577} and Peter M. Rost\, \orcidlink{0000-0002-8341-6989}}
\IEEEauthorblockA{Communications Engineering Lab (CEL), Karlsruhe Institute of Technology (KIT), Hertzstr.~16, 76187 Karlsruhe, Germany}
Email: \texttt{peter.gu@student.kit.edu, \{johannes.voigt, peter.rost\}@kit.edu}
}
\maketitle

\begin{abstract}
Due to an ever-increasing number of participants and new areas of application, the demands on mobile communications systems are continually increasing.
In order to deliver higher data rates, enable mobility and guarantee QoS requirements of subscribers, these systems and the protocols used are becoming more complex.
By using higher frequency spectrums, cells become smaller and more base stations have to be deployed.
This leads to an increased number of handovers of user equipments between base stations in order to enable mobility, resulting in potentially more frequent radio link failures and rate reduction.
The persistent switching between the same base stations, commonly referred to as ``ping-pong'', leads to a consistent reduction of data rates.
In this work, we propose a method for handover optimization by using proximal policy optimization in mobile communications to learn an adaptive handover protocol.
The resulting agent is highly flexible regarding different travelling speeds of user equipments, while outperforming the standard 5G NR handover protocol by 3GPP in terms of average data rate and number of radio link failures.
Furthermore, the design of the proposed environment demonstrates remarkable accuracy, ensuring a fair comparison with the standard 3GPP protocol.
\end{abstract}%

\begin{IEEEkeywords}
Communication Protocols, Handover, Mobility Management, Deep Reinforcement Learning
\end{IEEEkeywords}

\IEEEpeerreviewmaketitle

\section{Introduction}
The increasing number of users of wireless cellular networks and \ac{IoT} devices requires continuous adaptation and further development of mobile communication standards.
Furthermore, the system is expected to perform reliably, flexibly, fast, and with low latency in order to meet the increasing demands \cite{ab14}.
Current mobile standards are approaching channel capacity, and an increase in data rate is only possible through an expansion of the spectrum towards higher frequencies.
These frequencies experience significantly larger free-space path loss, leading to a substantial reduction in cell size and necessitating the deployment of numerous small \acp{BS}.
This leads to the problem that an \ac{UE}, e.g., a smartphone, has to connect to more \acp{BS} while moving.
If the \ac{UE} moves between two \acp{BS}, the so-called \ac{HO} process, which involves switching between \acp{BS}, can be redundantly executed.
This effect is called \textit{\ac{PP}}.
Finding the ideal time steps to trigger \acp{HO} is an optimization problem due to the data rate being zero during \ac{HO} processes.

In LTE and 5G NR, this challenge is tackled with a \ac{HO} protocol, standardized by \ac{3GPP}.
This protocol uses fixed parameters, e.g. \ac{TTT}, which limits the flexibility of it.
However, there are several approaches trying to optimize parameters with \ac{ML} models, e.g. \cite{hp18}.
In addition to that, work has been done to avoid those fixed parameters using \ac{RL}. Authors in \cite{yrh20} and \cite{lkw21} use a \ac{RL} model with discretized states, for example connection power, which lead to inaccuracies for the learning agent. This issue can be solved using \ac{DRL}, which is proposed in \cite{cllc21}.
However, the training process is done on the \acp{UE}, which will reduce the battery life of the device drastically.
Other approaches either use low \ac{UE} speeds \cite{mkk21} or \ac{SINR} instead of \ac{RSRP} values as \ac{NN} input, although the decision process of the standardized protocol uses \ac{RSRP} values.
Guo et al. also use the \ac{PPO} algorithm  \cite{gtzl20}, but without a direct penalization of \acp{PP} and \acp{HOF} as we did in our work.
Furthermore, there is no work that models the \ac{HO} process properly to guarantee better comparability with the \ac{3GPP} protocol. Our work is capable of modelling the \ac{HO} process in an accurate way, considering \ac{HO} preparation and execution period, while the model is well comparable with the \ac{3GPP} protocol since only \ac{RSRP} is used for training.  
In contrast to prior research that mainly concentrates on the \ac{HO} problem itself, we further introduce the capability of \ac{RL} models to operate at different \ac{UE} speeds while outperforming the \ac{3GPP} protocol. This is all done with no additional training for different speeds. 
The source code is available at GitHub\footnote{The source code of this work is publicly available at: \url{https://github.com/kit-cel/HandoverOptimDRL}}.

\section{preliminaries}
\subsection{Handover Procedure in 5G NR}
Similar to previous \ac{3GPP} releases, mobility in 5G NR is enabled by a mobility management function.
The decision, as to which \ac{BS} an \ac{UE} is connected to or to which \ac{BS} a handover is performed as soon as the \ac{UE} leaves the serving cell, is made using an event-based protocol and is based on \ac{RSRP} measurements.
Depending on the serving cell \ac{RSRP} and the \ac{RSRP} of neighbouring cells, different events A1--A6 will be triggered \cite[pp. 85 -- 89]{etsi18}.
Event A1 indicates that the \ac{RSRP} of the serving \ac{BS} is higher than a certain threshold, while event A2 is triggered if the \ac{RSRP} is lower, respectively.
The condition for a \ac{HO} is satisfied when the \ac{RSRP} of the serving \ac{BS} falls below an acceptable threshold (event A2) and at the same time the \ac{RSRP} of a neighbour \ac{BS} exceeds the \ac{RSRP} of the serving \ac{BS} by a given hysteresis value (event A3).
After a certain waiting time, \ac{TTT}, the \ac{UE} starts to transmit measurement reports to the serving \ac{BS}, in which, e.g., \ac{RSRP} measurements from various cells are reported.
Based on these reports, a \ac{HO} can be initiated by the serving \ac{BS}.
The hysteresis, cell-dependent offsets and \ac{TTT} are adjustable parameters and are used to avoid \acp{RLF} and reduce \acp{PP}.
If the serving \ac{BS} becomes too weak and a \ac{HO} is initiated too late, \ac{HOF} can occur, resulting in \acp{RLF}.
A \ac{RLF} means that the \ac{UE} is no longer synchronized with the network due to insufficient \ac{SINR}.
A \ac{HOF} occurs when a \ac{HO} process is aborted because the \ac{SINR} of the serving or to-be-served \acp{BS} have been too low for an extended period.

\subsection{Reinforcement Learning}\label{ch:RL}
In \ac{RL}, an agent interacts with an environment and learns a strategy (policy) on which its actions are based on.
The agent's actions are based on observations and states of the environment.
For actions taken, the agent receives rewards from the environment on which the policy learning is based, where the agent's goal is to maximize both immediate and future rewards \cite[pp. 1-5]{sb20}.

\begin{figure}[h]
    \centering
	\tikzset{
      frame/.style={
        rectangle, draw,
        text width=6em, text centered,
        minimum height=4em,fill=white,
        rounded corners,
      },
      line/.style={
        draw, -{latex},rounded corners=3mm,
      }
    }
	
	\begin{tikzpicture}[node distance = 4cm]
    \node [frame] (agent) {Agent};
    \node [frame, below=1.2cm of agent] (environment) {Environment};
    \draw[line] (agent) -- ++ (3.5,0) |- (environment) 
    node[right,pos=0.25,align=left] {$a_t$};
    \coordinate[left=8mm of environment] (P);
    \draw[thin,dashed] (P|-environment.north) -- (P|-environment.south);
    \draw[line] (environment.200) -- (P |- environment.200)
    node[midway,above]{$s_{t+1}$};
    \draw[line] (environment.160) -- (P |- environment.160)
    node[midway,above]{$r_{t+1}$};
    \draw[line] (P |- environment.200) -- ++ (-1.4,0) |- (agent.160)
    node[left, pos=0.25, align=right] {$s_t$};
    \draw[line] (P |- environment.160) -- ++ (-0.8,0) |- (agent.200)
    node[right,pos=0.25,align=left] {$r_t$};
\end{tikzpicture}
\caption{Schematic of \ac{RL}.}
\label{rl}
\end{figure}
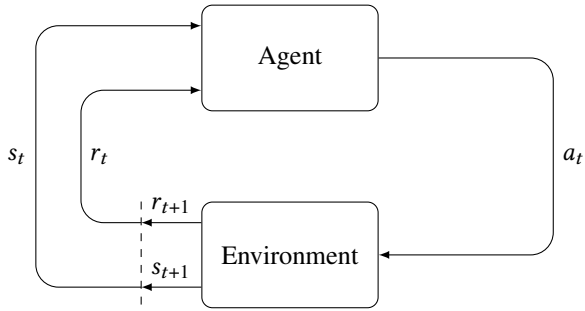

The following notation is used to describe the \ac{RL} system: state $s\in\mathcal{S}, s'=s_{t+1}$, action $a\in\mathcal{A}$, transition $(s, a, s')$, the transition probability distribution $p(s'|s, a)$ and the reward $r(s, a) = r_t$ in time step $t$.
In Fig.~\ref{rl}, the interaction between the agent and its environment is illustrated. The agent performs an action $a_{t}$ in its environment. Subsequently, a new state $s_{t+1}$ is reached and the agent receives a reward $r_{t+1}$ for its selected action.

The agent follows a \textit{policy} $\pi(a|s)$ and its goal is to maximize the expected future reward $\mathbb{E}_{\pi} \left[\sum_{k=0}^{\infty}\gamma^kr_{t+k}\right]$ with $\gamma\in[0,1]$ as the discount factor.
To evaluate the agent and thus the policy, the reward must be maximized. This reward is represented using the so-called \textit{V-function} \cite[p. 59]{sb20} 
{\small \begin{align}V^{\pi}_t(s) = \mathbb{E}_{\pi} \left[\sum_{k=0}^{\infty}\gamma^{k} r_{t+k}|s_t = s\right]  = \mathbb{E}_{\pi}\left[r_t + \sum_{k=1}^{\infty}\gamma^k r_{t+k}|s_t = s\right]\end{align}.}
 
Besides several \ac{RL} algorithms, there do exist algorithms which optimizations are based on the policy $\pi$. In \ac{PPO}, two \acp{NN}, called actor net and critic net, are used. The actor net is responsible for action decision, while the critic net is involved in the calculation of the loss function.

With $\theta$ as the weights of the \textit{actor net}, the calculation aims to select the policy that executes the optimal action $a^{\star}$ such that 
\begin{align}
    \theta_{t+1} = \theta_t + \alpha \nabla_{\theta} \pi_{\theta_t}(a^{\star}|s),
\end{align}
where $\alpha \in \mathbb{R}$ is the learning rate. After normalization and gradient weighting, this will lead to the commonly used estimator\footnote{This derivation step is not trivial. A good explanation can be found in \cite{eco18}.} \cite{eco18, swd17}
\begin{align}
	\hat{g} = \hat{\mathbb{E}}_t\left[\nabla_{\theta}\log\pi_{\theta}(a_t|s_t)\hat{A}_t\right],
    \label{eq:estimator}
\end{align}
with the advantage function $\hat{A}_t = \hat{A}(s_t|a_t)= Q^{\pi}(s_t, a_t) - V^{\pi}(s_t)$ \cite{swd17}. For calculating the V-values, the  \textit{critic net} with its weights $\xi$ is used.
As stated in \cite{swd17}, the probability ratio $\psi_t(\theta) = \frac{\pi_{\theta}(a_t|s_t)}{\pi_{\theta_{\text{old}}}(a_t|s_t)}$ is used for the estimator stated in Eq.~\ref{eq:estimator}. Next, authors in \cite{swd17} introduce the surrogate loss function used for \ac{PPO} 
\begin{align}
    L^{\text{CLIP}} = \hat{\mathbb{E}}_t \left[\text{min}\left(\psi_t(\theta)\hat{A}_t, ~\text{clip}(\psi_t(\theta), 1-\epsilon, 1+\epsilon)\hat{A}_t\right)\right].
    \label{eq:lossfunc}
\end{align}
This loss function can prevent excessively large update steps. By using $\epsilon\in (0,1)$ in this equation, the new policy $\pi_{\theta}$ is constrained to not deviate too far from the old policy $\pi_{\theta_{\text{old}}}$.

Now the actor-critic update process is roughly described. In the first step, the agent explores trajectories and store the resulting actions, rewards, and states into memory $D$, which can be obtained in Fig.~\ref{fig:ppo_diagram}. In the second step, V-values are delivered by the critic net while the net itself is updated via some gradient descent algorithm $L_2$, e.g. \ac{MSE}. In the third step, the actor net can be updated with those V-values, following the objective $L^{\text{CLIP}}$ \cite{oai}. 
\tikzstyle{process} = [rectangle, 
		minimum width=2.5cm, 
		minimum height=1cm, 
		text centered, 
		text width=2.5cm, 
		draw=black,
		]
		\tikzstyle{process2} = [rectangle, 
		minimum width=1cm, 
		minimum height=1cm, 
		text centered, 
		text width=1cm, 
		draw=black, 
		]
		\tikzstyle{circle-green} = [circle, 
		minimum size=0.2cm,
		text centered, 
 		text width=0.4cm, 
		draw=black,
		]
        \tikzstyle{circle-green2} = [circle, 
		minimum size=0.5cm,
		text centered, 
 		text width=0.8cm, 
		draw=black,
		]
		\tikzstyle{circle-tiny} = [circle,  
		inner sep=2pt, outer sep=0pt,
		text centered, 
		draw=black, 
		fill,
		]
		\tikzstyle{arrow} = [->,>=latex]
        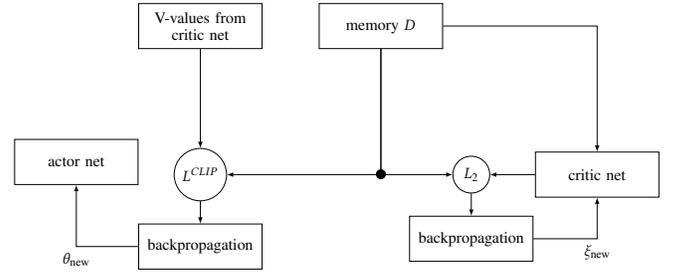
\begin{figure}
        \centering
        \scalebox{0.6}{
            \begin{tikzpicture}[node distance=4cm]
				
				\node (start) [process] {memory $D$};
				\node (oldv) [process, left of = start] {V-values from critic net};
				\node (actor) [process, below left = 2cm and 4cm of  start] {actor net};
				\node (l1) [circle-green2, below = 2.2cm of oldv] {$L^{CLIP}$};
				\node (l2) [circle-green, right = 5cm of l1]{$L_2$};
				\node (bp2) [process, below = 0.5cm of l2]{backpropagation};
				\node (bp1) [process, below = 0.5cm of l1]{backpropagation};
				\node (critic) [process, right = 1cm of l2]{critic net};
                \node (tinycirc) [circle-tiny, below = 2.68cm of start]{};
				
				\draw [arrow] (oldv) -- (l1);
				\draw [arrow] (l1) -- (bp1);
				\draw [arrow] (bp1) -| node[anchor=north] {$\theta_{\text{new}}$} (actor);
				\draw [arrow] (start) |- (l1);
				\draw [arrow] (start) |- (l2);
				\draw [arrow] (l2) -- (bp2);
				\draw [arrow] (critic) -- (l2);
				\draw [arrow] (bp2) -| node[anchor=north] {$\xi_{\text{new}}$} (critic);
				\draw [arrow] (start) -| (critic);
			\end{tikzpicture}}
        \caption{Flow diagram visualizing the actor-critic update process.}
        \label{fig:ppo_diagram}
        \end{figure}

\section{Handover Optimization Problem}
To achieve the highest data rate, one possibility is to immediately switch to the strongest \ac{BS}. However, as explained earlier, the data rate is zero during the \ac{HO} execution, which must be taken into consideration. Thus, a low number of \acp{PP} and \acp{HOF} is desirable. Furthermore, the optimal time for triggering a \ac{HO} has to be identified to prevent \acp{RLF}.

\section{System Model}\label{ch:system_model}
\begin{figure}
    \centering
    \includegraphics[scale=0.2]{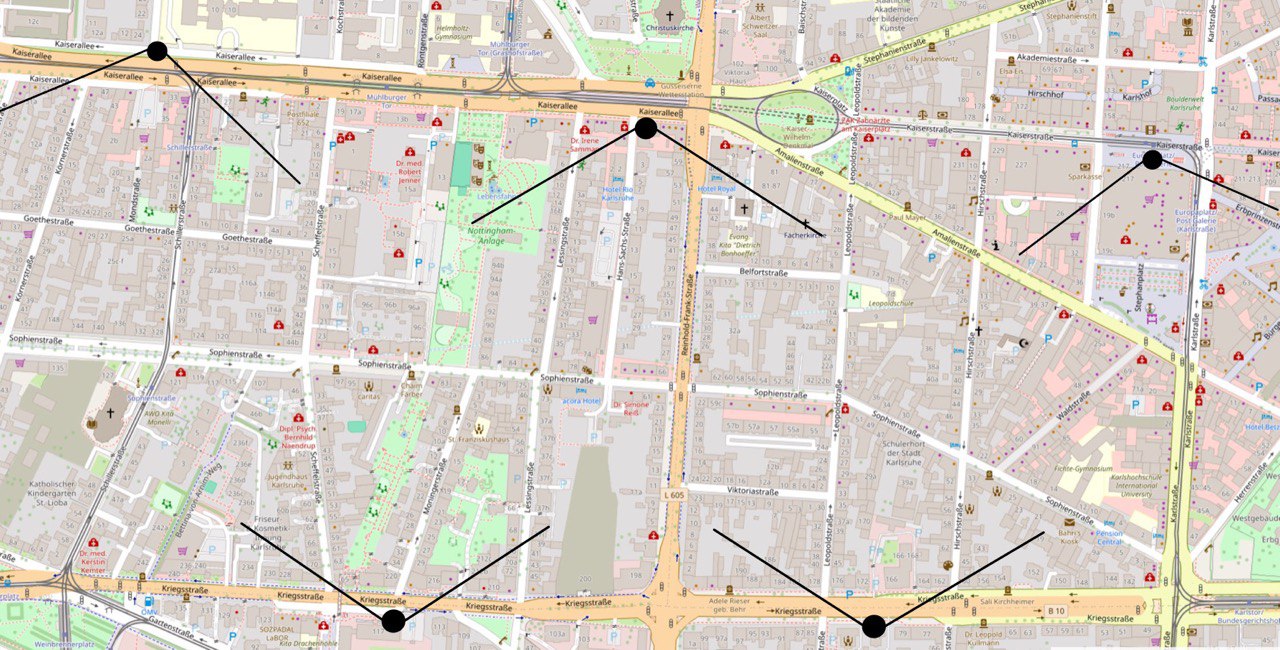} 
    \caption{Simulation area of Karlsruhe with 5 \acp{BS}. The black circles represent the \ac{BS}, while the lines indicate the direction of the beam.}
    \label{fig:karlsruhe_map}
\end{figure}

For the generation of \ac{RSRP} and \ac{SINR} traces of an \ac{UE} moving through an environment, the \textit{Vienna 5G System Level Simulator} \cite{vienna5g} is used.
An area of downtown Karlsruhe\footnote{Coordinates: 4\ang 900'17.6"N \ang 822'38.3"E -- 49°00'40.3"N 8°23'44.2"E} is used for simulations in this work. Overall, five 5G NR macro \acp{BS} are distributed over the map, as it is shown in Fig.~\ref{fig:karlsruhe_map}.
Furthermore, the \textit{QuaDRiGa} \cite{quadriga} channel model is used to simulate the link between an \ac{UE} and the base stations.
Information about the environment, such as floor plans of buildings and streets, is obtained via OpenStreetMap.
\acp{UE} are modelled such that they move only on streets at different speeds and thereby representing different user groups, e.g., pedestrians, cyclists and inner-city car traffic.
The simulated \ac{RSRP} and \ac{SINR} values are validated by comparing the distributions of the generated data with the distributions of measurement data in an urban area based on publicly available measurements of the 5G campus network in Kaiserslautern \cite{kaiserslautern}.

\subsection{\ac{RL} Environment Design}
\subsubsection{Simulation of Ping-Pongs}    
For simulation purposes, \ac{3GPP} suggests a minimum connection duration of an \ac{UE} with a \ac{BS}, referred to as the \ac{MTS} \cite{3gpp12}.
When a second \ac{HO} occurs within a time interval shorter than \ac{MTS} to the former \ac{BS}, it is referred to as \ac{PP}.

\subsubsection{Simulation of Handover Failures}\label{ch:sim_ho}
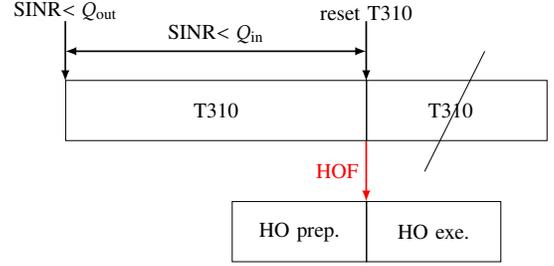
\begin{figure}
\centering
	\tikzstyle{process} = [rectangle, 
	minimum width=2cm, 
	minimum height=1cm, 
	text centered, 
	text width=2cm, 
	draw=black, 
	]
	\tikzstyle{process2} = [rectangle, 
	minimum width=0cm, 
	minimum height=0cm, 
	text centered, 
	text width=0cm, 
	draw=white, 
	]
	\tikzstyle{process3} = [rectangle, 
	minimum width=5cm, 
	minimum height=1cm, 
	text centered, 
	text width=4cm, 
	draw=black, 
	]
	\tikzstyle{process4} = [rectangle, 
	minimum width=3cm, 
	minimum height=1cm, 
	text width=2cm, 
	draw=black, 
	]
	\tikzstyle{arrow} = [draw=red,thick,->,>=latex]
	\tikzstyle{arrow2} = [draw=black,thick,->,>=latex]	
	\tikzstyle{arrow3} = [draw=black,thick,<->,>=latex]

\scalebox{.8}{\begin{tikzpicture}[node distance=2cm]
		\node (start) [process2]{};
		\node (in1) [process, right of=start, xshift=1.62cm] {HO prep.};
		\node (pro1) [process, right of=in1, xshift=0.23cm] {HO exe.};
		\node (optimize) [process3, above = 1cm of in1, xshift=-1.385cm] {T310};
		\node (optimize2) [process4, right of =optimize, xshift=2cm] {\hspace{1.5em}T310};
		\draw [arrow] (4.73,1.5) -- node[anchor=east] {\textcolor{red}{HOF}}(4.73,0.5);
		\draw[black] (5.7,1) to (6.7,3);
		\draw [arrow2] (4.73,3.5) -- node[yshift = 1.1em, anchor=south] {reset T310}(4.73,2.5);
		\draw [arrow2] (-0.27,3.5) -- node[yshift = 1.1em, anchor=south] {SINR$<Q_{\text{out}}$}(-0.27,2.5);
		\draw [arrow3] (-0.27,3) -- node[anchor=south] {SINR$<Q_{\text{in}}$}(4.73,3);
	\end{tikzpicture}}
\caption{\ac{HOF} triggered after \ac{HO} preparation.}
\label{fig:hof}
\end{figure}

To simulate \ac{HOF}, \cite{3gpp12} refers to the following method:
to detect a \ac{HOF}, a monitoring process runs concurrently beside the \ac{HO} procedure.
If the \ac{SINR} is worse than a configured threshold $Q_{\text{out}}$, indicating an unsynchronized communication link, timer T310 is initiated.
Two situations may arise:
\begin{itemize}
    \item If the \ac{HO} preparation is completed, and a handover is executed while the timer is running, a \ac{HOF} is triggered as shown in Fig.~\ref{fig:hof}.
    \item If timer T310 expires before a \ac{HO} can be executed, a \ac{HOF} is triggered.
\end{itemize}

Timer T310 can be stopped and reset if the \ac{SINR} exceeds a configured threshold $Q_{\text{in}}$ during this operation.
When a \ac{HOF} is triggered, the \ac{UE} undergoes a \ac{RLF} recovery process lasting multiple time steps.
During this period, the \ac{UE} is not connected to any \ac{BS}.
It establishes a new connection to a \ac{BS}, where this phase is called \ac{RLF} recovery.

\subsection{Simulation Overall}
To ensure a fair comparison with the \ac{HO} protocol of \ac{3GPP}, its rules and constraints, as explained earlier, have been met and consolidated.
The resulting algorithm is illustrated in Alg.~\ref{alg:system}.

\subsection{Pseudo RSRQ}
As mentioned before, only \ac{RSRP} values will be utilized in the \ac{HO} decision process to ensure high compatibility with the existing \ac{3GPP} protocol. To further address issues related to interference power, a \textit{pseudo \ac{RSRQ}} is used. The \ac{RSRQ} is defined in \cite{ash15} as:
\begin{align}
    \mathrm{RSRQ} = N \cdot \frac{\mathrm{RSRP}}{\mathrm{RSSI}},
    \label{eq:rsrq}
\end{align}
where $N$ is the number of resource blocks used and \ac{RSSI} is the average-received power in a single \ac{OFDM} symbol.
The measurement also includes the power of neighbouring \acp{BS}. %
Thus, this equation can be simplified as the ratio of the \ac{RSRP} of the serving \ac{BS} to the sum of \ac{RSRP} values from all interfering \acp{BS}.
If there exist $B = |\mathcal{B}|$ \acp{BS}, with indices $b = 0, 1,\ldots, B-1$, Eq.~\ref{eq:rsrq} can be simplified to:
\begin{align*}
    \mathrm{RSRQ}_{\mathrm{pseudo}, \mathrm{b}} = \frac{\mathrm{RSRP}_b}{\sum\limits_{i=0, i\neq b}^{B-1} \mathrm{RSRP}_i}.
\end{align*}
For simplicity, $\mathrm{RSRQ}_{\mathrm{pseudo}}$ will be abbreviated as $\mathrm{RSRQ}$ in the following sections.

\begin{algorithm}[h]
	\caption{}
	\begin{algorithmic}[1]
        \If{\ac{RLF} recovery not in progress}
            \If{T310 in progress}
                \If{T310 expired}
                    \State{Trigger \ac{HOF}}
                \Else
                    \If{\ac{HO} execution pending}
                        \State{Trigger \ac{HOF}}
                    \Else
                        \If{$\ac{SINR}>Q_{\mathrm{in}}$}
                            \State{Reset T310}
                        \EndIf
                    \EndIf
                \EndIf
            \Else
                \If{\ac{HO} over}
                    \If{\ac{SINR} of new \ac{BS} $< Q_{\mathrm{out}}$}
                        \State{Trigger \ac{HOF}}
                    \EndIf
                    \Else
                        \If{\ac{SINR} $< Q_{\mathrm{out}}$}
                            \State{Start T310}
                        \Else
                            \State{Reset T310}
                        \EndIf
                \EndIf
            \EndIf
        \EndIf
	\end{algorithmic}
	\label{alg:system}
\end{algorithm}

\subsection{State and Action Design}
The state $s\in \mathcal{S} = [\vec{S}_{\text{one-hot}}, \overrightarrow{\mathrm{SINR}}_{\text{norm}}, \vec{S}_{\mathrm{add}}]$ consists of three parts. The first component is a one-hot encoding vector $\vec{S}_{\text{one-hot}}$, indicating to which \ac{BS} the \ac{UE} is currently connected. All bits in the binary vector are set to `0', except for one bit which is set to `1' \cite{hg20}, e.g. $\vec{S}_{\text{one-hot}, b= 1} = [0,1,0]$ with $B=3$. 

The second part is composed of normalized and clipped \ac{RSRQ} values of all \acp{BS} in one time step. \ac{RSRQ} is clipped between \SI{-10}{dB} and \SI{10}{dB} to strengthen the focus on usable link quality and to eliminate ambiguities in finding the strongest \ac{BS}. These values are described as:
\begin{align*}
	\mathrm{RSRQ}_{\text{norm}} = 
	\begin{cases}
		1, &\mathrm{if ~RSRQ} \geq \mathrm{\SI{10}{dB}} \\
		0, &\mathrm{if ~RSRQ} \leq \mathrm{\SI{-10}{dB}} \\
		\frac{\mathrm{RSRQ} +10}{20}, &\mathrm{else.} \\
	\end{cases}	
\end{align*}
The clipping and the normalization are done to improve the learning performance of the \ac{NN} \cite{coh22}.

The third part $\vec{S}_{\mathrm{add}}$ is introduced to inform the \ac{NN} whether a \ac{PP} is possible. This is accomplished with 
\begin{align*}
	\vec{S}_{\mathrm{add}} = 
	\begin{cases}
		1, & \text{if $t- t_{\text{HO}}<$ MTS}\\
		0, & \text{else},
	\end{cases}
\end{align*}
with the current time step $t$ and the time step of the last \ac{HO} $t_{\text{HO}}$. This allows the agent to better identify whether a \ac{PP} or a \ac{HOF} is responsible for a negative reward. The state $s$ is the input of the \ac{NN}.

The actions $a\in \mathcal{A} = \{0, 1, \ldots, B-1\}$ serve as indicators of the preferred \ac{BS} to which the agent aims to establish a connection. This is obtained from the output of the \ac{NN}.

\subsection{Reward Design}
If the \ac{UE} runs into a \ac{RLF} recovery, the agent should be penalized by a constant $C\in \mathbb{R}^{+}$, as well as if \acp{PP} are detected or the \ac{SINR} of the \ac{BS} is less than $Q_{\text{out}}$.
Avoiding situations where $\mathrm{SINR} < Q_{\text{out}}$ is important to get rid of too late \ac{HO}, in which the \ac{SINR} becomes too low, triggering timer T310 and resulting in an \ac{RLF}.
Moreover, the time needed for \ac{RLF}-recovery is significantly longer than the \ac{HO} execution time.
So its penalty is twice as high as in the case of a \ac{HO}.
Upon successful \ac{HO}, the reward $r$ defined in Eq.~\ref{eq:reward_ppo} is increased by $C$ to acknowledge a successful \ac{HO}.
\resizebox{\linewidth}{!}{
\begin{minipage}{\linewidth}
\begin{align}
	r = 
	\begin{cases}
        \mathrm{RSRQ_{norm}} + C, & \text{\ac{UE} connected to the strongest \ac{BS}}\\
		  \mathrm{RSRQ_{norm}} , &\begin{aligned}
		& \text{\ac{UE} is connected to a strong, but not} \\ & \text{the strongest \ac{BS}}
        \end{aligned}\\
        - C, &\begin{aligned}
        & \text{\ac{UE} is connected to a \ac{BS} with} \\ &\text{\ac{SINR}$< Q_{\text{out}}$}
        \end{aligned}\\
		- C, & \text{\ac{PP} detected} \\
		- 2\cdot C, & \text{\ac{UE} is not connected, \ac{RLF} recovery}\\
	\end{cases} \label{eq:reward_ppo}
\end{align}
\vspace*{0.01pt}
\end{minipage}
}

\subsection{Training Process}
In each epoch, a random dataset, which is uniformly distributed, will be chosen for training. Each data set consists of both \ac{SINR} and \ac{RSRP} values of all \acp{BS}. While \ac{SINR} is used only for monitoring purposes (see Ch.~\ref{ch:sim_ho}), it is not utilized for \ac{RL} training. To enhance diversity in the data, \ac{BS}--\ac{RSRP}\slash\ac{SINR} mappings are randomly selected in every epoch during training. Since actions cannot be executed during \ac{HO} execution or \ac{RLF} recovery, those time steps will be skipped. During training, the agent gets terminated if either \acp{HOF} or \acp{PP} occur. The following chapter provides a more detailed explanation of the learning algorithm.

\subsection{Learning Algorithm}
In the first step, weights of actor and critic net have to be initialized. 
After selecting a data set and shuffling it, the agent will explore various trajectories, generating samples as it selects actions and store them in a memory $D$ \cite{oai}\cite{pa18}. If this is done, the objectives explained in Ch. ~\ref{ch:RL} are executed. The exploration will be terminated in case of \acp{HOF} or \acp{PP}. As a consequence, the agent must begin exploring the dataset from time step $0$.
The learning algorithm of the model can be found in Alg.~\ref{alg:ppo}. 

\begin{algorithm}[h]
	\caption{}
	\begin{algorithmic}[1]
		\State{$\theta \leftarrow$ Initialize weights of actor net}
		\State{$\xi \leftarrow$ Initialize weights of critic net}
		\For{episode $i$}
        \State{Choose one dataset and shuffle the \ac{RSRP}\slash\ac{SINR} \hspace*{1.2em} mapping}
		\State\parbox[t]{\dimexpr\linewidth-\algorithmicindent}{The agent explores trajectories of the environment, generates $m$ samples ${(s, a, s', r)}$, and stores them in the memory $D$}
		\For{1 to $m/n$}
		\State\parbox[t]{\dimexpr\linewidth-\algorithmicindent}{Choose a data set batch consisting of $n$ samples}
		\State{Calculate V values of the critic net}
		\For{each sample with its corresponding V value}
		\State{Calculate the V value using the current critic \hspace*{4.2em} net to get $\hat{A}$.}
		\State{$\theta \leftarrow \theta_{\text{new}}$ with $L^{\text{CLIP}}$}
		\State{$\xi \leftarrow \xi_{\text{new}}$}
        \If{HOF or PP occurs}
        \State{Termination -- reset environment to time \hspace*{5.7em} step 0}
        \EndIf
		\EndFor 
		\EndFor
		\State{Clear the memory $D$}
		\EndFor
	\end{algorithmic}
	\label{alg:ppo}
\end{algorithm}

\section{Performance Evaluation}
\subsection{Experimental Setup}
After generating \ac{SINR} and \ac{RSRP} values using the Vienna 5G Simulator \cite{vienna5g} for 15 different trajectories in the area described in Ch.~\ref{ch:system_model}, the \ac{PPO} agent will undergo training, utilizing Python and \ac{SB3}~library\footnote{Source code has been adjusted to disable mini-batch shuffling to stabilize training} \cite{rhg21}.
$10$ routes are for training and $5$ for testing.
The \acp{UE} moves at a speed of \SI{50}{km/h} for exactly 3~minutes each.
With a sample interval of \SI{120}{ms}, this will result in $1500$~time steps. \SI{120}{ms} has been chosen to align with the report interval standardized by \ac{3GPP} \cite{etsi18}.
The duration of certain events, such as the preparation/execution of \acp{HO}, is shorter than the sample interval.
In order to correctly represent these events, the samples generated using the Vienna 5G Simulator are interpolated such that the temporal resolution is \SI{10}{ms}.
This interpolation is achieved using the Fourier method.
To mitigate Fourier side effects, a moving average filter with a length of \SI{2}{s} is applied to the data.

For both, actor and critic network, the same network architecture of the hidden layers is used.
While the first and third hidden layers consist of $64$ neurons, the second hidden layer comprises $128$ neurons.
The ReLU activation function is employed in both the input layer and all hidden layers, whereas no activation function is utilized in the output layer.

The training process consists of three training iterations.
First, the agent will randomly choose the entropy coefficient $\mathrm{ent}_{\mathrm{coef}}\in \{0.1, 0.01, 0.001\}$ and the constant $C \in [0.6, 0.95]$.
With those two parameters, the first iteration is executed on all datasets. To accelerate the learning process, all data sets are reduced to \SI{1}{min} for the initial and subsequent learning phase.
Furthermore, the environment will only be reset if a \ac{HOF} occurs.
The learning rate is set to $5\cdot 10^{-5}$ for the first $500$ epochs.
In the second iteration, the learning rate is set to $10^{-6}$ for further $300$ epochs.
In the last iteration the training will be executed on the whole dataset of \SI{3}{min} with the same learning rate and number of epochs as in iteration $2$.
Additionally, the environment will now also be reset if a \ac{PP} occurs.
In all three iterations, the learning rate decreases linearly until it reaches zero after the last epoch.

The evaluation of the learned agent is conducted on five traces of an \ac{UE} moving through the streets of the simulated area for each of the considered velocities: \SI{3}{km/h}, \SI{30}{km/h}, and \SI{50}{km/h}.
Finally, the agent is trained with a \ac{UE} moving at \SI{50}{km/h}, but is also evaluated at speeds of \SI{30}{km/h} and \SI{3}{km/h} without additional training required.

In the next section, the performance of the \ac{PPO} agent is evaluated and compared to the existing 5G NR \ac{HO} protocol implemented using Python 3.10 and following Alg.~\ref{alg:system}.
Three metrics are tracked: the mean data rate and the number of \acp{HOF} and \acp{PP}.
The mean data rate is compared to the mean of the maximum data rate at each time step $i$. This involves taking the logarithm of the respective maximum $\ac{SINR}$ at each time step and subsequently computing the overall average \cite[p. 13]{ps08}.
After obtaining the ratio of the \ac{SINR} values between the connected \ac{BS} and the \ac{BS} with the maximum \ac{SINR}, the mean data rate is calculated as follows:
\begin{align*}
	\Gamma
    &=
    \frac{\sum_{i=0}^{\Lambda-1} \log_2\left(1+\mathrm{SINR}_i\right)}{\sum_{i=0}^{\Lambda-1} \max_{b\in\mathcal{B}}\log_2\left(1+\mathrm{SINR}_{i,b}\right)},
\end{align*}
with the total number of time steps $\Lambda$.

\subsection{Simulation Results}
We compare the trained \ac{PPO} agent with the standard \ac{3GPP} protocol.
The parameters used by the \ac{3GPP} protocol are illustrated in Tab.~\ref{tab:3gpp}.
For an accurate model of the \ac{3GPP} handover procedure, the \ac{HO} preparation and execution time, timer T310, $Q_{\mathrm{in}}$, $Q_{\mathrm{out}}$, \ac{RLF} recovery and \ac{MTS} are also used in the environment of the \ac{PPO} setup.
Further parameters, tuned with \ac{wb} \cite{wandb} can be obtained from Tab. \ref{tab:ppo}.

\begin{table}[htp]
    \caption{Parameters used for \ac{3GPP} protocol \cite{hp18}\cite{3gpp12}.}
    \centering
    \fontsize{10pt}{14pt}\selectfont
    \begin{tabular}{{l}{c}}
    \hline
    Parameter & Value \\
    \hline 
    A2 threshold & \SI{-80}{dBm} \\
    A2 hysteresis & \SI{1}{dB} \\
    A3 hysteresis & \SI{1}{dB} \\
    \ac{HO} preparation time & \SI{50}{ms} \\
    \ac{HO} execution time & \SI{40}{ms} \\
    T310 & \SI{1000}{ms} \\
    \ac{TTT} & \SI{160}{ms} \\
    $Q_{\mathrm{out}}$ & \SI{-8}{dB} \\
    $Q_{\mathrm{in}}$ & \SI{-6}{dB} \\
    \ac{RLF} recovery & \SI{200}{ms} \\
    \ac{MTS} & \SI{1000}{ms} \\
    Offset & \SI{2}{dB} \\
    \hline
    \end{tabular}
    \label{tab:3gpp}
\end{table}

\begin{table}[htp]
    \caption{\ac{PPO} agent related parameters.}
    \centering
    \fontsize{10pt}{14pt}\selectfont
    \begin{tabular}{{l}{c}}
        \hline
        Parameter & Value  \\
        \hline
        $\mathrm{ent}_{\mathrm{coef}}$& 0.1 \\
        C & 0.9405\\
        batch size iteration 1 and 2 & 150 \\
        batch size iteration 3 & 550 \\
        \hline
    \end{tabular}
    \label{tab:ppo}
\end{table}

\begin{figure}[h]
\subfloat[\ac{3GPP} protocol]{\begin{tikzpicture}
        \begin{axis}[
            axis y line=left,
            xlabel = speed UE (\SI{}{km/h}),
            ylabel = numbers PP/HOF,
            ymin = 0,
            ymax = 8,
            xmin = 0,
            xmax = 50,
            ]
            \addplot[mark=square, blue] coordinates {(3,0) (30,3) (50,7)};
            \label{HOF}
            \addplot[mark=*, magenta] coordinates {(3,0) (30,0) (50,0)};
            \label{PP}
        \end{axis}
        \begin{axis}[
            axis y line=right,
            ylabel = $\Gamma$ (\%),
            ymin = 98,
            ymax = 100,
            xmin = 0,
            xmax = 50,
            legend style={at={(0.012,0.6)}, anchor=north west},
            ]
            \addlegendimage{/pgfplots/refstyle=HOF}\addlegendentry{HOF}
            \addlegendimage{/pgfplots/refstyle=PP}\addlegendentry{PP}
            \addplot[mark=triangle, black!60!green] coordinates {(3,98.51) (30,99.63) (50,99.35)};
            \label{datarate}
            \addlegendentry{$\Gamma$}
        \end{axis}
    \end{tikzpicture}
\label{fig:res_3gpp}}
\hfill
\subfloat[\ac{PPO} agent]{\begin{tikzpicture}
        \begin{axis}[
            axis y line=left,
            xlabel = speed UE (\SI{}{km/h}),
            ylabel = numbers PP/HOF,
            ymin = 0,
            ymax = 8,
            xmin = 0,
            xmax = 50,
            ]
            \addplot[mark=square, blue] coordinates {(3,0) (30,0) (50,0)};
            \label{HOF}
            \addplot[mark=*, magenta] coordinates {(3,1) (30,3) (50,0)};
            \label{PP}
        \end{axis}
        \begin{axis}[
            axis y line=right,
            ylabel = $\Gamma$ (\%),
            ymin = 98,
            ymax = 100,
            xmin = 0,
            xmax = 50,
            legend style={at={(0.012,0.6)}, anchor=north west},
            ]
            \addlegendimage{/pgfplots/refstyle=HOF}\addlegendentry{HOF}
            \addlegendimage{/pgfplots/refstyle=PP}\addlegendentry{PP}
            \addplot[mark=triangle, black!60!green] coordinates {(3,99.84) (30,99.89) (50,99.94)};
            \label{datarate}
            \addlegendentry{$\Gamma$}
        \end{axis}
    \end{tikzpicture}%
\label{fig:res_ppo}}
\caption{Simulation results for the \ac{3GPP} protocol and the \ac{PPO} agent for one evaluation data set.}
\label{fig:res}
\end{figure}
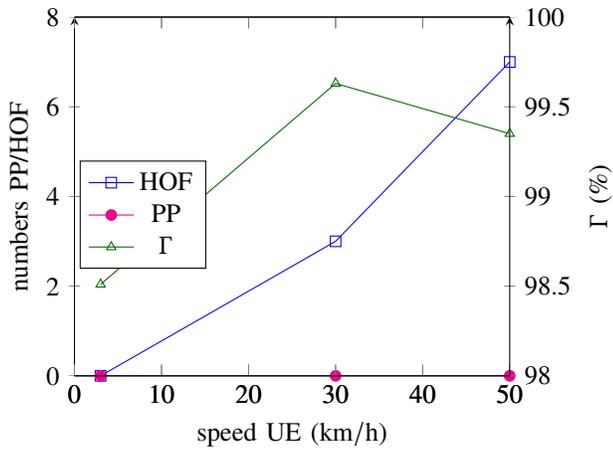
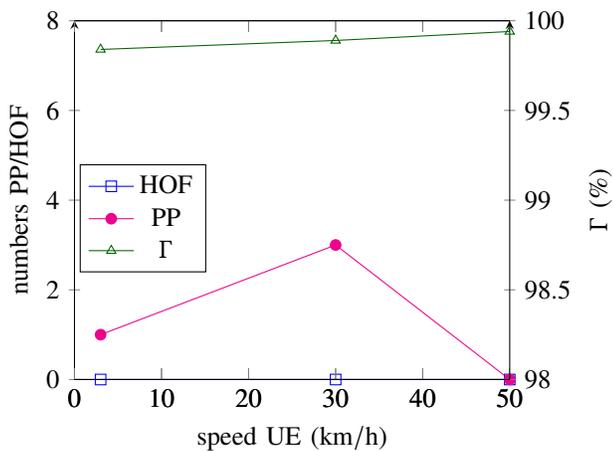

Figure~\ref{fig:res} shows the simulation results of one evaluation dataset used for both the \ac{3GPP} protocol and the \ac{PPO} agent.
The three metrics \ac{HOF}, \ac{PP} and $\Gamma$ are plotted for three different \ac{UE} speeds.
On this dataset, the \ac{3GPP} protocol does not have any problems with \acp{PP} but the number of \acp{HOF} increases with higher \ac{UE} speeds.
Compared to the \ac{3GPP} protocol, the \ac{PPO} agent is able to trigger the \acp{HO} such that no \acp{HOF} occurs.
However, some \acp{PP} occur in this case.
It must be noted that the definition of a \ac{PP} is not unambiguous, and therefore the number of \acp{PP} occurred depends on the \ac{MTS} defined in \cite{3gpp12}.
Moreover, the mean data rate achieved by the \ac{PPO} agent consistently exceeds that of the \ac{3GPP} protocol. This trend is reflected in the metric $\Gamma$, which is consistently higher for the \ac{PPO} agent.

\section{Conclusion}
In this paper, we investigated a new method for learning adaptive \ac{HO} protocols by using \ac{PPO}.
Our results show that a \ac{RL}-based \ac{HO} protocol can outperform the standard protocol by \ac{3GPP} in terms of mean data rate and  number of \acp{RLF}.
The adaptivity of the \ac{RL} agent to varying environment conditions is remarkable, e.g., different user velocities, without changing any parameters or additional training for different conditions.
The \ac{HO} decisions of the agent are based on \ac{RSRP} measurements without using further information about the simulated area or the positions of the \ac{UE} and the \acp{BS}, to ensure a fair comparison to the \ac{3GPP} protocol.
For the comparison of our model to the standard protocol, we implemented an abstraction of the \ac{3GPP} standard, including the \ac{HO} preparation and  execution time as well as timers for \ac{RLF} detection.

\bibliographystyle{IEEEtran}
\bibliography{references-short}

\begin{thebibliography}{10}
\providecommand{\url}[1]{#1}
\csname url@samestyle\endcsname
\providecommand{\newblock}{\relax}
\providecommand{\bibinfo}[2]{#2}
\providecommand{\BIBentrySTDinterwordspacing}{\spaceskip=0pt\relax}
\providecommand{\BIBentryALTinterwordstretchfactor}{4}
\providecommand{\BIBentryALTinterwordspacing}{\spaceskip=\fontdimen2\font plus
\BIBentryALTinterwordstretchfactor\fontdimen3\font minus
  \fontdimen4\font\relax}
\providecommand{\BIBforeignlanguage}[2]{{%
\expandafter\ifx\csname l@#1\endcsname\relax
\typeout{** WARNING: IEEEtran.bst: No hyphenation pattern has been}%
\typeout{** loaded for the language `#1'. Using the pattern for}%
\typeout{** the default language instead.}%
\else
\language=\csname l@#1\endcsname
\fi
#2}}
\providecommand{\BIBdecl}{\relax}
\BIBdecl

\bibitem{ab14}
J.~G. Andrews, S.~Buzzi, W.~Choi, S.~V. Hanly, A.~Lozano, A.~C.~K. Soong, and
  J.~C. Zhang, ``What {Will} {5G} {Be}{?}'' \emph{IEEE Journal on Selected
  Areas in Communications}, vol.~32, no.~6, pp. 1065--1082, Jun. 2014.

\bibitem{hp18}
D.~Castro-Hernandez and R.~Paranjape, ``Optimization of {Handover} {Parameters}
  for {LTE}/{LTE}-{A} in-{Building} {Systems},'' \emph{IEEE Transactions on
  Vehicular Technology}, vol.~67, no.~6, pp. 5260--5273, Jun. 2018.

\bibitem{yrh20}
V.~Yajnanarayana, H.~Rydén, and L.~Hévizi, ``{5G} {Handover} using
  {Reinforcement} {Learning},'' in \emph{2020 {IEEE} 3rd {5G} {World} {Forum}
  ({5GWF})}, Sep. 2020, pp. 349--354.

\bibitem{lkw21}
Q.~Liu, C.~F. Kwong, S.~Wei, L.~Li, and S.~Zhang,
  ``\BIBforeignlanguage{en}{Intelligent {Handover} {Triggering} {Mechanism} in
  {5G} {Ultra}-{Dense} {Networks} {Via} {Clustering}-{Based} {Reinforcement}
  {Learning}},'' \emph{\BIBforeignlanguage{en}{Mobile Networks and
  Applications}}, vol.~26, no.~1, pp. 27--39, Feb. 2021.

\bibitem{cllc21}
Y.~Cao, S.-Y. Lien, Y.-C. Liang, and K.-C. Chen, ``Federated {Deep}
  {Reinforcement} {Learning} for {User} {Access} {Control} in {Open} {Radio}
  {Access} {Networks},'' in \emph{{ICC} 2021 - {IEEE} {International}
  {Conference} on {Communications}}, Jun. 2021, pp. 1--6.

\bibitem{mkk21}
M.~Mollel, S.~Kaijage, and K.~Michael, ``\BIBforeignlanguage{en}{Deep
  {Reinforcement} {Learning} based {Handover} {Management} for {Millimeter}
  {Wave} {Communication}},'' 2021.

\bibitem{gtzl20}
D.~Guo, L.~Tang, X.~Zhang, and Y.-C. Liang, ``Joint {Optimization} of
  {Handover} {Control} and {Power} {Allocation} {Based} on {Multi}-{Agent}
  {Deep} {Reinforcement} {Learning},'' \emph{IEEE Transactions on Vehicular
  Technology}, vol.~69, no.~11, pp. 13\,124--13\,138, Nov. 2020.

\bibitem{etsi18}
\BIBentryALTinterwordspacing
``Specification\hspace{3pt}\#:\hspace{3pt}38.331.'' [Online]. Available:
  \url{https://portal.3gpp.org/desktopmodules/Specifications/SpecificationDetails.aspx?specificationId=3197}
\BIBentrySTDinterwordspacing

\bibitem{sb20}
R.~S. Sutton and A.~G. Barto, \emph{Reinforcement Learning: An Introduction},
  2nd~ed.\hskip 1em plus 0.5em minus 0.4em\relax The MIT Press, 2018.

\bibitem{eco18}
\BIBentryALTinterwordspacing
A.~L. Ecoffet. (2018) An {Intuitive} {Explanation} of {Policy} {Gradient}.
  [Online]. Available:
  \url{https://towardsdatascience.com/an-intuitive-explanation-of-policy-gradient-part-1-reinforce-aa4392cbfd3c}
\BIBentrySTDinterwordspacing

\bibitem{swd17}
\BIBentryALTinterwordspacing
J.~Schulman, F.~Wolski, P.~Dhariwal, A.~Radford, and O.~Klimov, ``Proximal
  {Policy} {Optimization} {Algorithms},'' Aug. 2017, arXiv:1707.06347 [cs].
  [Online]. Available: \url{http://arxiv.org/abs/1707.06347}
\BIBentrySTDinterwordspacing

\bibitem{oai}
\BIBentryALTinterwordspacing
OpenAI, ``Openai spinning up - proximal policy optimization,'' 2020. [Online].
  Available: \url{https://spinningup.openai.com/en/latest/algorithms/ppo.html}
\BIBentrySTDinterwordspacing

\bibitem{vienna5g}
M.~K. Müller, F.~Ademaj, T.~Dittrich, A.~Fastenbauer, B.~Ramos~Elbal,
  A.~Nabavi, L.~Nagel, S.~Schwarz, and M.~Rupp, ``Flexible multi-node
  simulation of cellular mobile communications: the {Vienna} {5G} {System}
  {Level} {Simulator},'' \emph{EURASIP Journal on Wireless Communications and
  Networking}, vol. 2018, no.~1, p. 227, Sep. 2018.

\bibitem{quadriga}
\BIBentryALTinterwordspacing
``\BIBforeignlanguage{en}{{QuaDRiGa}}.'' [Online]. Available:
  \url{https://quadriga-channel-model.de/}
\BIBentrySTDinterwordspacing

\bibitem{kaiserslautern}
S.~B. Mallikarjun, C.~Schellenberger, C.~Hobelsberger, and H.~D. Schotten,
  ``Performance {Analysis} of a {Private} {5G} {SA} {Campus} {Network},'' in
  \emph{Mobile {Communication} - {Technologies} and {Applications}; 26th
  {ITG}-{Symposium}}, May 2022, pp. 1--5.

\bibitem{3gpp12}
\BIBentryALTinterwordspacing
``Specification\hspace{3pt}\#:\hspace{3pt}36.839.'' [Online]. Available:
  \url{https://portal.3gpp.org/desktopmodules/Specifications/SpecificationDetails.aspx?specificationId=2540}
\BIBentrySTDinterwordspacing

\bibitem{ash15}
F.~Afroz, R.~Subramanian, R.~Heidary, K.~Sandrasegaran, and S.~Ahmed,
  ``\BIBforeignlanguage{en}{{SINR}, {RSRP}, {RSSI} and {RSRQ} {Measurements} in
  {Long} {Term} {Evolution} {Networks}},''
  \emph{\BIBforeignlanguage{en}{International Journal of Wireless \& Mobile
  Networks}}, vol.~7, no.~4, pp. 113--123, Aug. 2015.

\bibitem{hg20}
\BIBentryALTinterwordspacing
F.~Harrag and S.~Gueliani, ``Event extraction based on deep learning in food
  hazard arabic texts,'' \emph{CoRR}, vol. abs/2008.05014, 2020. [Online].
  Available: \url{https://arxiv.org/abs/2008.05014}
\BIBentrySTDinterwordspacing

\bibitem{coh22}
\BIBentryALTinterwordspacing
J.~Coholich, ``\BIBforeignlanguage{en-us}{A {Bag} of {Tricks} for {Deep}
  {Reinforcement} {Learning}},'' May 2022. [Online]. Available:
  \url{https://jmcoholich.github.io/post/rl_bag_of_tricks/}
\BIBentrySTDinterwordspacing

\bibitem{pa18}
X.~B. P., P.~Abbeel, S.~Levine, and M.~van~de Panne, ``Deepmimic:
  Example-guided deep reinforcement learning of physics-based character
  skills,'' 2018.

\bibitem{rhg21}
A.~Raffin, A.~Hill, A.~Gleave, A.~Kanervisto, M.~Ernestus, and N.~Dormann,
  ``Stable-baselines3: reliable reinforcement learning implementations,''
  \emph{The Journal of Machine Learning Research}, vol.~22, no.~1, pp.
  268:12\,348--268:12\,355, Jan. 2021.

\bibitem{ps08}
J.~G. Proakis and M.~Salehi, \emph{\BIBforeignlanguage{eng}{Digital
  communications}}, 5th~ed., ser. {McGraw}-{Hilll} higher education.\hskip 1em
  plus 0.5em minus 0.4em\relax Boston [u.a.]: McGraw-Hill, 2009.

\bibitem{wandb}
\BIBentryALTinterwordspacing
L.~Biewald, ``Experiment tracking with weights and biases,'' 2020, software
  available from wandb.com. [Online]. Available: \url{https://www.wandb.com/}
\BIBentrySTDinterwordspacing

\end{thebibliography}

\begin{acronym}
  \acro{3GPP}{3rd Generation Partnership Project}
  \acro{ANN}{Artificial Neural Network}
  \acro{BP}{Backpropagation}
  \acro{BS}{base station}
  \acro{CNN}{convolutional neural network}
  \acro{DQN}{deep Q-learning}
  \acro{DRL}{deep reinforcement learning}
  \acro{GAE}{generalized advantage estimation}
  \acro{GD}{gradient descent}
  \acro{HO}{handover}
  \acro{HOF}{handover failure}
  \acro{IoT}{Internet of things}
  \acro{lr}[Lr]{learning rate}
  \acro{lstm}[LSTM]{long short-term memory}
  \acro{MC}{Monte Carlo estimation}
  \acro{MDP}{Markov decision process}
  \acro{ML}{machine learning}
  \acro{MR}{measurement report}
  \acro{MSE}{mean squared error}
  \acro{MTS}{minimum time of stay}
  \acro{NN}{neural network}
  \acro{OFDM}{orthogonal frequency division multiplexing}
  \acro{PA}{polyak averaging}
  \acro{PP}{ping pong}
  \acro{PPO}{proximal policy optimization}
  \acro{RAN}{radio access network}
  \acro{RAT}{radio access technology}
  \acro{RB}{replay buffer}
  \acro{RL}{reinforcement earning}
  \acro{RLF}{radio link failure}
  \acro{RRC}{radio ressource control}
  \acro{RSRP}{reference signal received power}
  \acro{RSRQ}{reference signal received quality}
  \acro{RSSI}{received signal strength indicator}
  \acro{UE}{user equipment}
  \acro{SB3}{Stable Baselines3}
  \acro{SINR}{signal-to-interference-plus-noise ratio}
  \acro{SIR}{Signal-to-interference ratio}
  \acro{TTT}{time-to-trigger} 
  \acro{wb}[W\&B]{Weights \& Biases}
\end{acronym}

\end{document}